\documentclass[aps,pra,twocolumn]{revtex4-1}

\usepackage{graphics}
\usepackage{epsfig}

\usepackage{sidecap}
\usepackage{wrapfig}

\begin{document}


\title{A Hubbard model for ultracold bosonic atoms interacting via
zero-point-energy induced three-body interactions}

\author{Saurabh Paul}
\affiliation{Center for Quantum Information and Computer Science and
Joint Quantum Institute, University of Maryland, College Park, Maryland
20742, USA}

\author{P. R. Johnson}
\affiliation{Department of Physics, American University, Washington DC 20016, USA}

\author{Eite Tiesinga}
\affiliation{Joint Quantum Institute and Center for Quantum Information
and Computer Science, National Institute of Standards and Technology
and University of Maryland, Gaithersburg, Maryland 20899, USA}

\begin{abstract}
We show that for ultra-cold neutral bosonic atoms held in a
three-dimensional periodic potential or optical lattice, a Hubbard
model with dominant, attractive three-body interactions can
be generated. In fact, we derive that the effect of pair-wise
interactions can be made small or zero starting from the realization that
collisions occur at the zero-point energy of an optical lattice
site and the strength of the interactions is energy
dependent from effective-range contributions.  We determine the strength of
the two- and three-body interactions for scattering from van-der-Waals
potentials and near Fano-Feshbach resonances.  For van-der-Waals
potentials, which for example describe scattering of alkaline-earth atoms, we find
that the pair-wise interaction can only be turned off for species with
a small negative scattering length, leaving the $^{88}$Sr isotope a
possible candidate.  Interestingly, for collisional magnetic Feshbach
resonances this restriction does not apply and there often exist
magnetic fields where the two-body interaction is small.  We illustrate
this result for several known narrow resonances between alkali-metal atoms
as well as chromium atoms.  Finally, we compare the size of the three-body
interaction with hopping rates and describe limits due to three-body
recombination.  \end{abstract}

\maketitle

In 1998 Jaksch {\it et al.} \cite{Jaksch1998} suggested that 
laser-cooled atomic samples can be held in optical lattices, periodic potentials
created by counter-propagating laser beams.  These three-dimensional
lattices have spatial periods between 400 nm and 800 nm and depths $V_0$
as high as $V_0/h \sim 1$ MHz, where $h$ is Planck's constant. An ensemble
of atoms then realize either the fermionic or bosonic Hubbard model, where
atoms hop from site to site and interact only when on the same site.
The interaction driven quantum phase transition of this model was first
realized by Ref.~\cite{Greiner2002a}.

Today, optical lattices are seen as a natural choice in which to
simulate other many-body Hamiltonians.  These include Hamiltonians
with complex band structure such as double-well lattices
\cite{Lee2007,Qian2011,Olschlager2012,Soltan2012}, two-dimensional hexagonal
lattices \cite{Snoek2007,Soltan2012,Uehlinger2013,Jurgensen2014}, as well as those with
spin-momentum couplings possibly leading to topological matter
\cite{Lin2009,Lin2011}.  Quantum phase transitions in
these Hamiltonians enable ground-state wavefunctions with unusual
order parameters, such as pair superfluids and striped phases 
\cite{Hu2009,Bolukbasi2014,Brown2015}. Phase transitions in
Hamiltonians with long-range dipole-dipole interactions using atoms
or molecules with large magnetic or electric dipole moments can also be
studied. Finally, atoms in optical lattices can be used to measure
gravitational acceleration (little-$g$) \cite{Rosi2014,Mahmud2014a,Meinert2014}, shed light
on non-linear measurements \cite{Giovannetti2011,Javanainen2012,Tiesinga2013,Mahmud2014b}, and be used for quantum information processing.

Over the last ten years ultra-cold atom experiments have also
investigated few-body phenomena.  In particular, three-body interactions
have been studied through Efimov physics of strongly
interacting atoms observed as resonances in three-body recombination,
where three colliding atoms create a dimer and a free atom
\cite{Kraemer2006,Braaten2007,Ferlaino2010}.  Here, recent developments
include the prediction of a minimum in the recombination rate coefficient
$K_3$ for scattering of a van-der-Waals potential with a $d$-wave
shape resonance \cite{Wang2012}. Moreover, Ref.~\cite{Wang2014} presented
advanced numerical simulations that can quantitatively model observed
recombination rates,  while Ref.~\cite{Shotan2014} showed empirically
that for a broad $^7$Li Feshbach resonance, $K_3$ is controlled by the
effective range correction of the atom-atom scattering.

Proposals that suggest ways to create atomic gasses dominated by elastic
three-body interactions have also been made.  In Refs.~\cite{Petrov2014,Daley2014} this
was achieved by adding resonant radiation to couple internal states of
an atom or by driving the lattice at rf frequencies. Some of us showed
that the low-energy behavior of atoms in complex lattice geometries
(i.e. double-well optical lattices) can also be engineered to lead
to large three-body interactions \cite{Paul2015}.  
Interestingly, after the observation of the formation of droplets \cite{Kadau2015} in a ferromagnetic atomic dysprosium
condensate induced by a rapid quench to attractive pair-wise interactions Refs.~\cite{Xi2015,Bisset2015} have  
independently suggested  that the origin of this instability are large repulsive elastic three-body collisions

In this paper we propose a novel way to create dominant three-body
interactions in Hubbard models. We rely on two ingredients.  The first
relies on the analytical analysis of scattering from a van-der-Waals
potential \cite{Gao2009,Gribakin1993} as well as analytical modeling
of Fano-Feshbach resonances, where the energy of molecular states is
tuned with a magnetic-field \cite{Chin2010}.  This analysis confirms that
ultra-cold scattering is describable in terms of a scattering length $a$
and effective range $r_e$ that are uniquely specified by the van-der-Waals
coefficient and resonance parameters.  The second ingredient is the
realization that two-, three-, and higher-body interaction energies
of atoms in an optical lattice site can under certain assumptions be
computed analytically \cite{Johnson2009,Yin2014}.

We will show that for two atoms in a lattice site, with a non-negligible
zero-point energy, a cancellation of the contribution from the scattering
length and effective range contribution can occur while simultaneously
three atoms have a finite inseparable three-body interaction that is of
sufficient magnitude that an experimental observation is possible.

This paper is organized as follows. In section \ref{sec:review} we
introduce delta-function interactions between atoms with strength defined
by the scattering length and effective range and review results for the
ground-state energy of a few atoms held in a site of an optical lattice.
We also examine the quality of a harmonic approximation of the lattice
site potential.  In Sec.~\ref{sec:cancel} we derive the relationship
between $a$ and $r_e$ for which the two-body interactions cancel and three-body
interactions remain.
Sections \ref{sec:vdW} and \ref{sec:FB} describe how this relationship
can be met for a van-der-Waals potential and for Feshbach resonances,
respectively. For scattering from a van-der-Waals potential we show that
the $^{88}$Sr isotope is a promising candidate.  For Feshbach resonances
we work out four cases, one each for $^{23}$Na, $^{39}$K, $^{52}$Cr,
and $^{133}$Cs scattering.  We also compare the expected three-body
interaction energies with tunneling energies between lattice sites.
Section \ref{sec:recomb} describes two methods to determine lattice
parameters for which there are no on-site two-body interactions and
discusses limits set by three-body recombination.

\section{Pseudo-potential for low-energy collisions, optical lattices,
and effective field theory} \label{sec:review}

In 1957 K.~Huang \cite{Huang1957} showed that the low-energy scattering
of two neutral atoms of mass $m$ with an isotropic inter-atomic potential
can be modeled by the equivalent three-dimensional delta-function pseudo-potential
\begin{equation}
       V_{\rm pseudo}(\vec R)= 4\pi \frac{\hbar^2}{2\mu} (a - \frac{1}{2} r_e a^2 \nabla^2) \delta(\vec R)
    \frac{\partial}{\partial R} R \,,
       \label{pseuso}
\end{equation}
where $\vec R$ describes the separation and orientation of the
atom pair, $\nabla$ is the gradient operator for the relative
motion, $\mu=m/2$ is the reduced mass, and $\hbar=h/(2\pi)$. 
The scattering length $a$ and the
effective range $r_e$ parametrize the effect of the physical
interaction potential.  (This derivation was revisited in
Refs.~\cite{Derevianko2005,Idziaszek2006,Pricoupenko2006}.)
Crucial for this paper is that  $a$ and $r_e$ have a
simple relationship and can be tuned near Feshbach resonances.

Our atoms are held in a three-dimensional periodic potential created
by counter-propagating laser beams with wavevectors $k_L$. For simplicity we assume a 
cubic lattice with potential $V(\vec x)=V_0\sum_i\cos^2(k_Lx_i)$,
where $\vec x=(x_1,x_2,x_3)$ is the atomic location and $V_0$ is the lattice depth.
The potential has periodicity $\pi/k_L$ and a minimum in each unit cell
with harmonic frequency and single-atom oscillator length 
given by
\[
      \hbar\omega =  2\sqrt{V_0E_R} 
\ \ {\rm and} \ \ 
       \ell = \sqrt{\hbar/(m\omega)} = 1/(k_L\sqrt[4]{V_0/E_R})\,,
\]
respectively. Here $E_R=\hbar^2k_L^2/(2m)$ is the recoil energy.

\begin{figure}
	\includegraphics[width=0.155\textwidth,trim=0 0 0 0,clip]{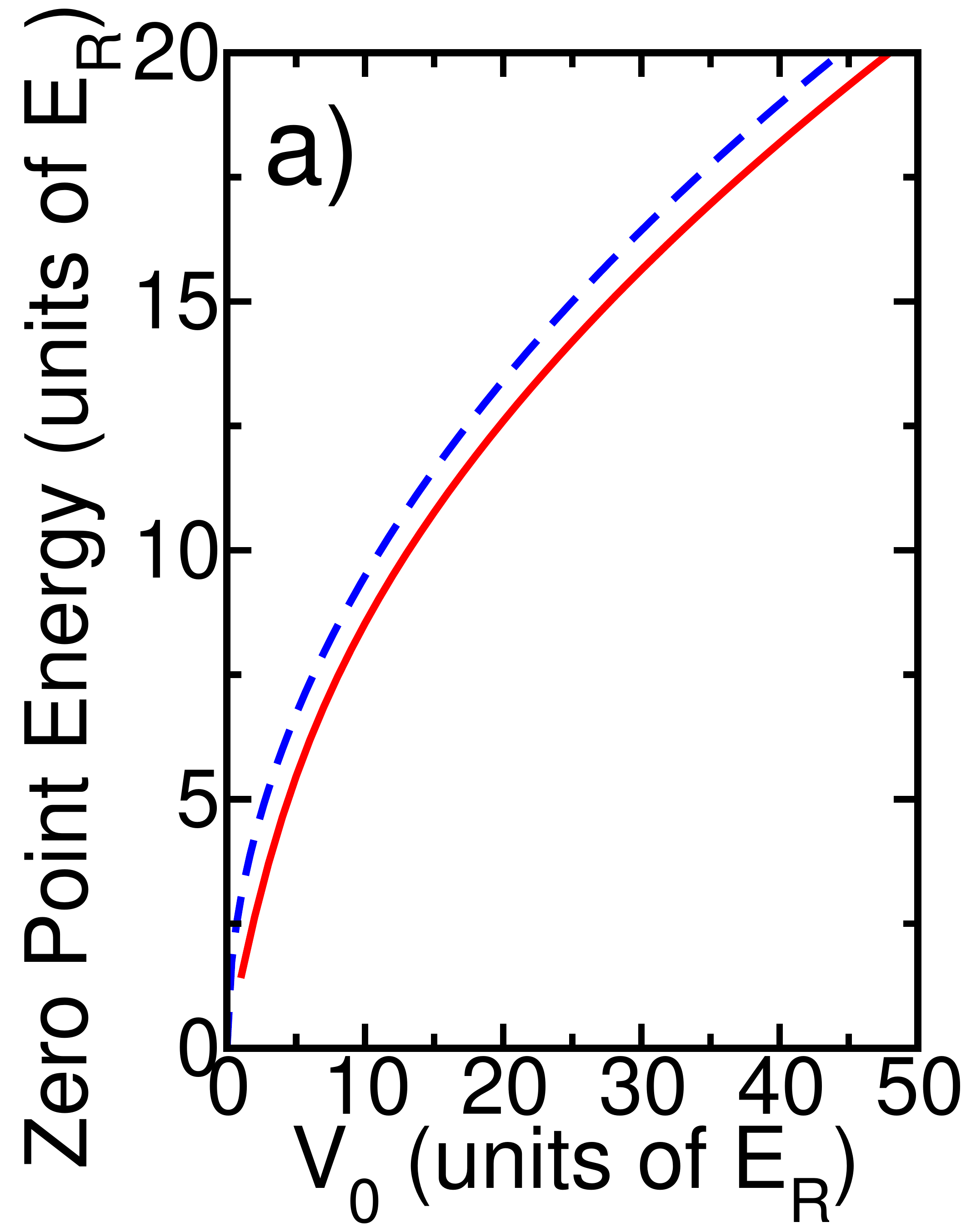}
\includegraphics[width=0.155\textwidth,trim=0 0 0 0,clip]{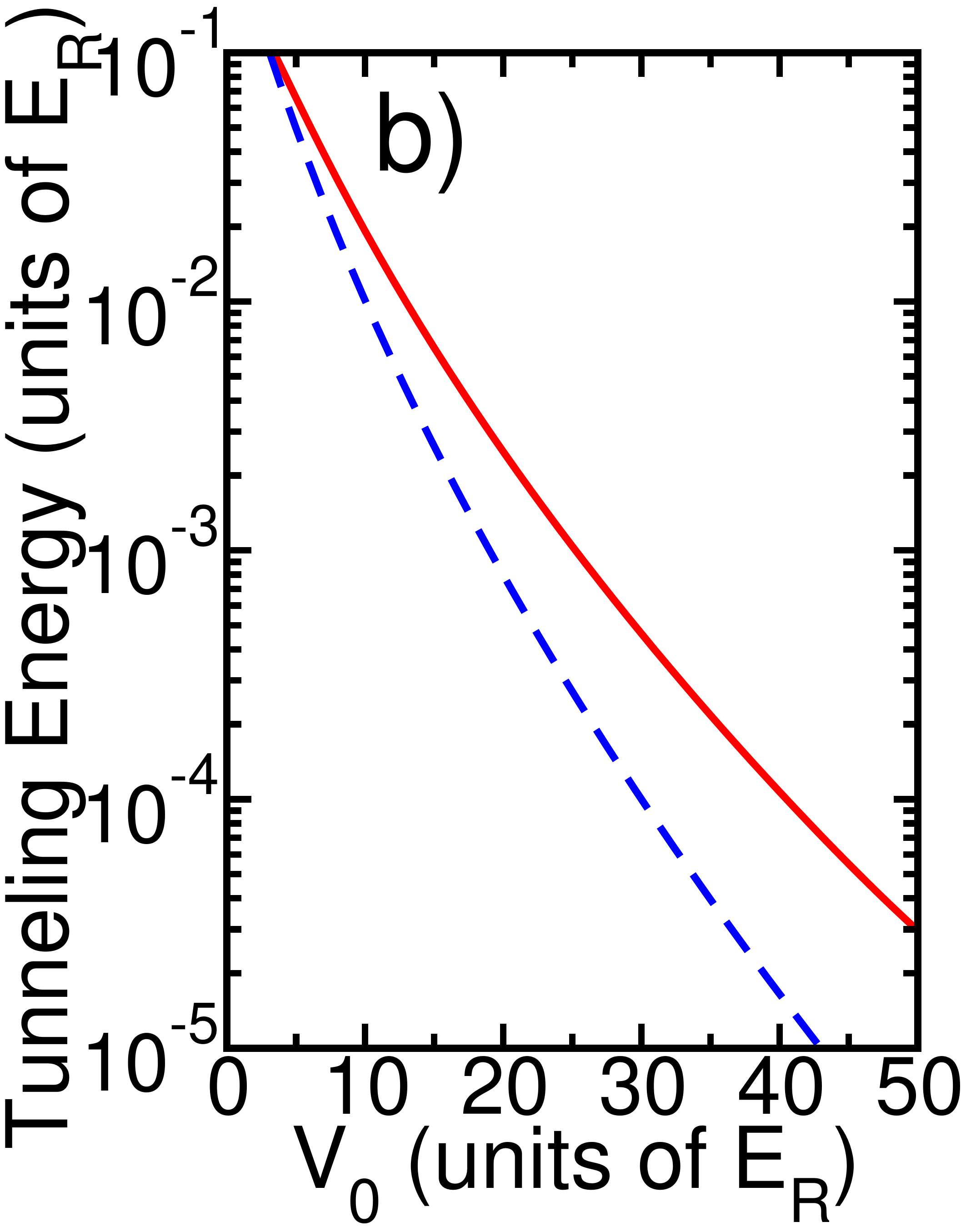}
\includegraphics[width=0.155\textwidth,trim=0 0 0 0,clip]{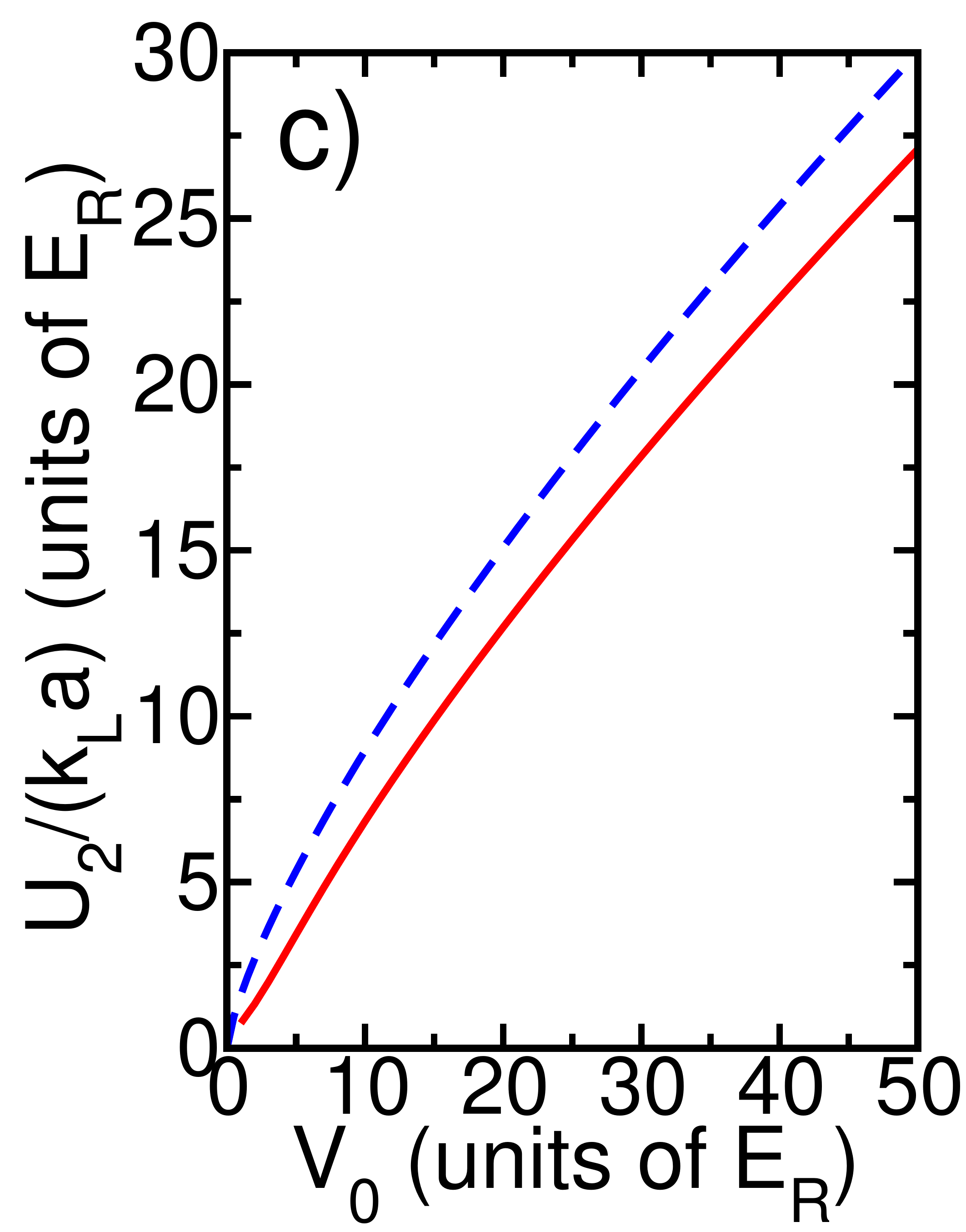}

\caption{(color online) Zero-point energy (panel a), tunneling energies
(panel b), and the scaled first-order two-body interaction strength
$U_2/(k_La)$ with no effective-range correction (panel c)
as a function of lattice depth $V_0$ for a cubic, three-dimensional
optical lattice.  Solid red curves are based on exact band-structure
calculations and exact Wannier functions.  Dashed blue lines are based
on oscillator solutions of the isotropic harmonic approximation around the
lattice minima \cite{vanOosten2001}. }
\label{fig:zpeJU}
\end{figure}

We will rely on this harmonic approximation near the lattice minima.
Figures~\ref{fig:zpeJU}(a) and (b) show that for sufficiently large
$V_0$ this is qualitatively correct. Panel (a) compares the zero point
energy of the harmonic approximation, $3\hbar\omega/2$, with that of the on-site energy
of the lowest band obtained from our exact band-structure calculation.
The exact on-site energy is always smaller since anharmonic corrections
are attractive.  Similarly, panel (b) shows a comparison of the tunneling
energies between nearest-neighbor unit cells.
Here, the perturbative (harmonic) result underestimates the tunneling energy because
anharmonic corrections delocalize the Wannier functions.

The harmonic approximation also simplifies the calculation of 
the interaction energies between atoms.  Non-perturbative eigenenergies
for two atoms interacting via a delta-function
potential were derived in Ref.~\cite{Busch1998}. Moreover, Refs.~\cite{Johnson2009,Yin2014} perturbatively calculated  the
ground-state energy $E_{n=2,3,\cdots}$ of two, three, or more atoms based on
effective-field theory \cite{Zee2010}. In fact, up to
second-order perturbation theory when $a\ll\ell$ and $r_ea^2/2\ll\ell^3$
they showed $E_n=3n\hbar\omega/2+U_2n(n-1)/2+U_3n(n-1)(n-2)/6$, where $U_2$
and $U_3$ are the two- and three-body interaction strengths
\begin{eqnarray}
   U_2/\hbar\omega &=&  \xi+\frac{3}{2}\epsilon +
         (1-\log 2) \xi^2
         + 2\left (2 - \frac{3}{2} \log 2  \right)  \xi\epsilon \nonumber\\
   && \quad
     + \left( \frac{15}{4} - \frac{9}{4}\log 2   \right) \epsilon^2\,,
\end{eqnarray}
and
\begin{eqnarray}
   U_3/\hbar\omega 
              &=&  \left\{6 - 4 \sqrt{3} -6 \log\left(\frac{4}{2+\sqrt{3}}\right) \right\} \xi^2 \label{eq:U3} \\
                && \quad
                     + \left\{24 - \frac{52}{3}\sqrt{3}  -18\log\left(\frac{4}{2+\sqrt{3}}\right)
                     \right\} \xi\epsilon \nonumber \\
                && \quad
                     + \left\{ \frac{45}{2} - \frac{55}{3}\sqrt{3}
                            -\frac{27}{2}\log\left(\frac{4}{2+\sqrt{3}} \right)
                       \right\} \epsilon^2  \,, \nonumber
\end{eqnarray}
with dimensionless quantities 
\[
 \xi=\sqrt{\frac{2}{\pi}} \frac{a}{\ell} \quad {\rm and}\quad 
 \epsilon= \sqrt{\frac{2}{\pi}} \frac{1}{2}\frac{r_ea^2}{\ell^3}\,,
\]
and $\log z$ is the natural logarithm. Four- and higher-body
interaction strengths are zero at this order of field theory.
Reference~\cite{Beane2007} performed similar calculations for a box with
periodic boundary conditions.  We ignore small corrections from non-zero
partial wave and anisotropic magnetic dipole-dipole scattering.

For completeness Fig.~\ref{fig:zpeJU}(c) compares the two-body interaction
strength in a harmonic trap evaluated to first-order in $a$ and $r_e=0$
(i.~e. $U_2=\sqrt{2/\pi} (a/\ell)\hbar\omega$) with the corresponding
matrix element based on the energetically-lowest Wannier function of the
three-dimensional optical lattice. The curves are in sufficiently good
agreement such that a harmonic approximation with its analytical results up
to second-order perturbation theory can be confidently used for the
analysis of $U_2$ and $U_3$. 

\section{Cancellation of the two-body interaction}
\label{sec:cancel}

We can now search for parameter regimes where $U_2$ is small
compared to $U_3$ and, in particular, look for the case $U_2=0$.
In fact, by factorizing $U_2$ and requiring
that $\xi\ll 1$ and $\epsilon\ll 1$
we realize that if we can achieve
\begin{equation}
      \epsilon=  -\frac{2}{3} \xi 
   \quad {\rm or}\quad
     \frac{1}{2}  r_e a^2=  -\frac{2}{3} a\ell^2
\label{eq:cancelU2}
\end{equation}
the two-body interaction strength $U_2$ vanishes as
the contributions from the scattering length and the
effective range cancel.  
Equation \ref{eq:cancelU2} can be shown to hold to all orders in $a$ and $r_e$
from Ref.~\cite{Busch1998} (by making the replacement $a\to a+ r_ea^2 (2\mu E/\hbar^2)/2$ in Eq.~16 of that article). 
More importantly, the three-body interaction strength does not vanish and is
\begin{equation}
   U_3/\hbar\omega = - \frac{16}{9} \frac{1}{\sqrt{3}} \xi^2 
   =- \frac{32}{9\pi\sqrt{3}}  \frac{a^2}{\ell^2} \,,
       \label{eq:cancelU3}
\end{equation}
which is always attractive and remains of the same order of magnitude
as in Eq.~\ref{eq:U3}. The next two sections describe ways in which we can
achieve this cancellation.

\section{van der Waals potential} \label{sec:vdW}

\begin{figure}
\includegraphics[width=0.48\textwidth,trim=0 0 0 0,clip]{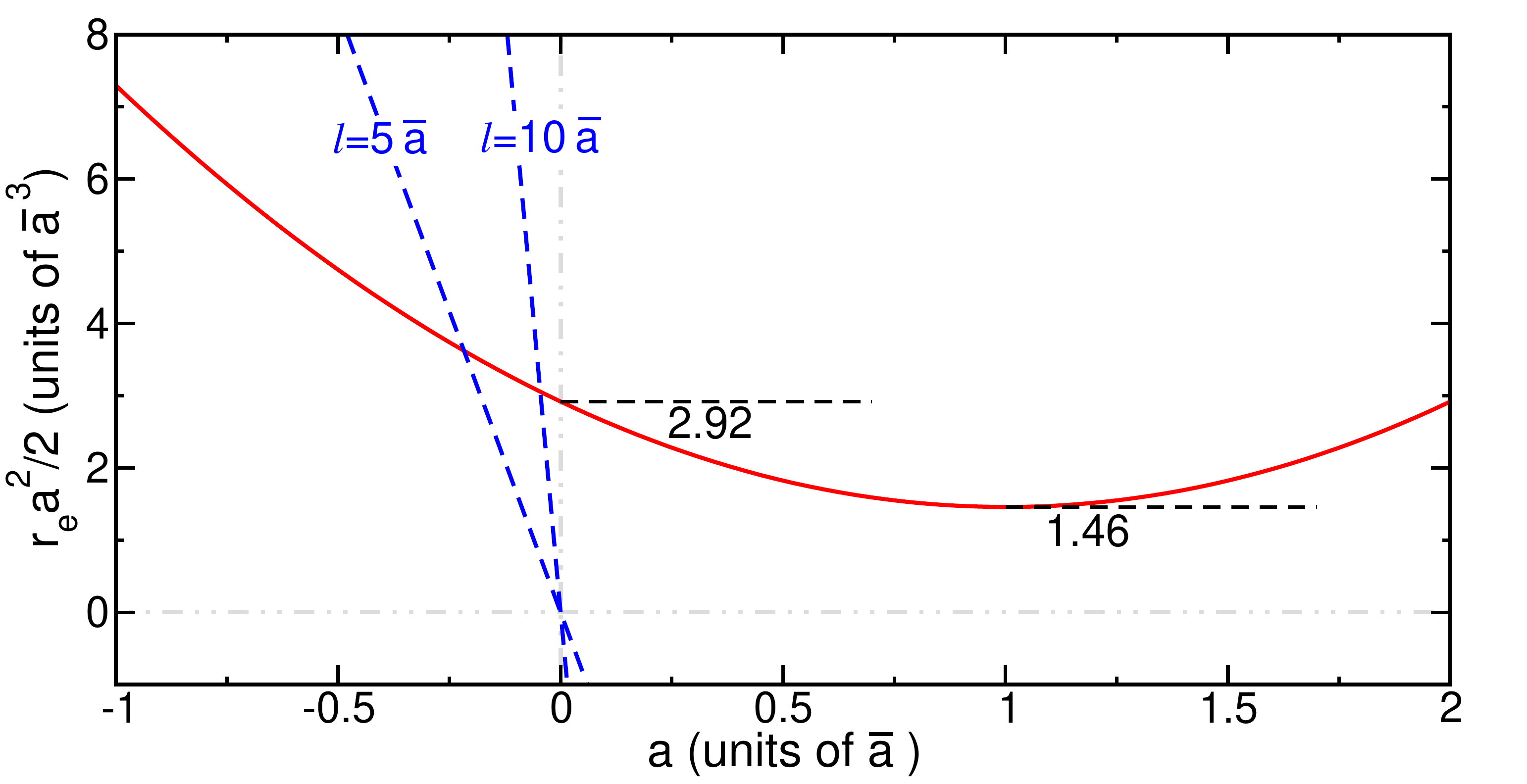}
\caption{(color online) The effective range volume $r_ea^2/2$ (solid red curve) as a function
of scattering length $a$ for a van der Waals potential. All lengths
are expressed in units of the mean scattering length ${\bar a}$. 
The dashed blue curves correspond to $-2 a\ell^2/3$ for two
values of $\ell$. At intersections of $r_ea^2/2$ and
$-2 a\ell^2/3$ the effective two-body interaction is tuned to zero.}
\label{vdW}
\end{figure}

Ultra-cold scattering between structureless ground-state atoms, such as
the alkaline-earth atoms, or between more-complex atoms away from any
scattering resonance, such as alkali-metal atoms in an external magnetic
field, is controlled by its long-range isotropic $-C_6/R^6$ potential,
where $C_6$ is the  van-der-Waals coefficient. This follows from the
fact that for separations where deviations from this van-der-Waals
potential due to electron bonding are significant, its depth is already
orders of magnitude larger than the initial kinetic energy of the
atoms \cite{Gao2009,Chin2010}.  References \cite{Flambaum1999,Gao1998} then showed that
when the potential has a scattering length $a$ its effective range is
\begin{eqnarray}
   \frac{1}{2} r_e a^2 &=&
                  \frac{1}{3c_e^2}  \bar a
                        \left( (a-\bar a)^2+{\bar a}^2 \right) \,,
    \label{eq:volumeVdW}
\end{eqnarray}
where ${\bar a} = c_e (2\mu C_6/\hbar^2)^{1/4}$ is the mean scattering
length \cite{Gribakin1993} and $c_e=2\pi/[\Gamma(1/4)]^2=0.4780 \cdots$,
and $\Gamma(z)$ is the Gamma function.  For typical atoms $\bar a$ lies
between $30a_0$ and $100a_0$, where $a_0=0.0529$ nm is the Bohr radius.
Figure~\ref{vdW} shows the effective range volume $r_ea^2/2$ as a
function of $a$. It is always positive, has a minimum at $a=\bar a$,
and for $a\to 0$ equals $r_ea^2/2=2.918\cdots {\bar a}^3$,
which implies that $r_e$ diverges for a zero scattering length.

In order to find regimes where $U_2$ is small compared to $U_3$,
we investigate whether Eq.~\ref{eq:cancelU2} can hold. This equality 
is graphically solved in Fig.~\ref{vdW} for two ratios $\ell/\bar a\gg 1$,
corresponding to typical circumstances in current experiments.
We immediately observe that solutions exist for negative scattering lengths
that are small compared to $\bar a$.
In fact, a Taylor expansion for large $\ell/{\bar a}$ gives
\begin{equation}
        \frac{a}{\bar a} = -   \frac{1}{c^2_e} \left(\frac{\bar a}{\ell}\right)^2 + O(1/\ell^4) 
     \label{eq:c1}
\end{equation}
and thus $|a/{\bar a}|\ll 1$ and $|a/\ell|   \ll 1$ consistent with our assumptions.

For a van-der-Waals potential $a$ is fixed.  Hence, Eq.~\ref{eq:c1}
is a constraint on $\ell$ or the trapping frequency $\omega$ (and thus
on the lattice depth $V_0$).  Moreover, there exist only a few atomic
species with the small negative scattering length needed to have a small
or vanishing $U_2$.  In fact, we are only aware of the strontium isotope
$^{88}$Sr to satisfy $|a/{\bar a}|\ll1$, since it has a scattering length
of $a=-2.0(3) a_0$ and $\bar a =71.76 a_0$ \cite{Stein2010}. (Numbers in
parenthesis are one-standard-deviation uncertainties.)  Hence, we find
that $U_2=0$ requires $\ell=900a_0$ and thus $\omega/(2\pi)=50$ kHz.
Assuming a realistic Sr optical lattice with a photon recoil energy of $E_R/h=4.0$
kHz, we read from Fig.~\ref{fig:zpeJU}a) that $V_0\approx 40E_R$ and
that from Fig.~\ref{fig:zpeJU}b) the tunneling energy $J\approx 10^{-4}
E_R$ or $J/h\approx 0.4$ Hz.  This tunneling energy is comparable
to the three-body strength $U_3/h\approx -0.15$ Hz calculated from
Eq.~\ref{eq:cancelU3}.


\section{Feshbach resonances} \label{sec:FB}

Ultracold scattering of alkali-metal atoms \cite{Chin2010} in a magnetic
field $B$ contains collisional resonances, where the scattering length
can be tuned.  Recently, interest has also focused on resonances with
atoms with large magnetic moments, such as Cr \cite{Werner2005}, Er
\cite{Frisch2014}, and Dy \cite{Petrov2012,Baumann2014}, as the long-range
magnetic dipole-dipole interaction influences their collective behavior.

At ultra-cold collision energies $E=\hbar^2k^2/(2\mu)$ resonant
scattering is described by the scattering amplitude
\cite{Fano1961,Taylor1972,Kohler2006}
\begin{eqnarray}
f(k)=f_{\rm bg}(k)- e^{2i\delta_{\rm bg}(k)}\frac{\Gamma(E)/2}{E-E_{\rm res}(B,E)+i\Gamma(E)/2} \,, 
 \label{eq:ff}
\end{eqnarray} 
where $f_{\rm bg}(k)=e^{i\delta_{\rm bg}(k)}\{\sin \delta_{\rm bg}(k)\}/k$
is the background scattering amplitude away from the resonance and
$\delta_{\rm bg}(k)$ is the background phase shift. We assume that
the low-energy behavior of $f_{\rm bg}(k)$ is that of a
van-der-Waals potential with scattering length $a_{\rm bg}$ as discussed
in Sec.~\ref{sec:vdW}.  The dispersive second term of Eq.~\ref{eq:ff}
describes the resonance with a magnetic-field and energy-dependent
resonance location $E_{\rm res}(B,E)=\mu_e(B-B_0)+\beta E$ and
positive energy width $\Gamma(E)=2(ka_{\rm bg}) \Gamma_0\times(1+\alpha
E/\Gamma_0)$, where $\mu_e$ is the magnetic moment of the resonant state,
$B_0$ is the magnetic field at resonance, and $\Gamma_0$ is the resonance
strength.  Finally, the field-independent coefficients $\alpha$ and
$\beta$ describe additional energy dependencies of $\Gamma(E)$ and $E_{\rm
res}(E)$ and will affect the effective range.

We note that by definition
$ {\rm Re}    f(k) =-  a - \left\{r_e a^2/2-a^3\right\} k^2 + \cdots $
and a Taylor expansion of Eq.~\ref{eq:ff} in $k$ 
then leads to the scattering length 
$ a= a_{\rm bg}-a_{\rm bg}\Gamma_0/E_{\rm res}(B,0) $
and  effective range volume 
\begin{eqnarray}
  \lefteqn{ \frac{1}{2}r_e a^2 = \frac{1}{2}r_{\rm bg}a_{\rm bg}^2
        + aa_{\rm bg}(a- a_{\rm bg}) \label{eq:direct}} \\
 && 
          -  \, (1-\beta)( a- a_{\rm bg}  )^2 {\bar a}/s_{\rm res}
  + \, \alpha(a-a_{\rm bg}) {\bar a}a_{\rm bg}/s_{\rm res}  \nonumber \\
 &&\ \ \ \quad \equiv\ V_{\rm q} + g_{\rm q} ( a- a_{\rm q} )^2 \,,
         \label{eq:volumeFR}
\end{eqnarray}
where $r_{\rm bg}$ is the background effective range given in
Eq.~\ref{eq:volumeVdW} when evaluated at scattering length $a_{\rm bg}$.
We have eliminated the dependence on $E_{\rm res}(B,0)$ in favor of
$a$ and the dimensionless $s_{\rm res}\equiv a_{\rm bg}\Gamma_0/({\bar
a}{\bar E})>0$ characterizes the resonance strength in terms of
the mean scattering length $\bar a$ and energy ${\bar E}=\hbar^2/(2\mu {\bar
a}^2)$ of a van-der-Waals potential \cite{Chin2010}.  A resonance is
narrow when $s_{\rm res}\ll 1$ and broad otherwise.
Moreover, the volume $r_e a^2/2
\to  r_{\rm bg}a_{\rm bg}^2/2$ when $a\to a_{\rm bg}$ 
as expected and
\[
  \frac{1}{2}r_e a^2 \to \frac{1}{2}r_{\rm bg}a_{\rm bg}^2 
  - (1-\beta+\alpha) {\bar a} a_{\rm bg}^2/s_{\rm res}  \,
\]
for $a\to 0$ showing that $r_e a^2/2$ can be negative. For narrow
resonances this was already noted in Ref.~\cite{Petrov2004}.

\begin{table}[t]
\caption{Parameters for five Feshbach resonances.  Columns represent the
atomic species, $B_0$ in Gauss, the background scattering length $a_{\rm
bg}$, resonance strength $s_{\rm res}$, coefficients $V_{\rm q}$, $g_{\rm
q}$ and $a_{\rm q}$, where available from Ref.~\cite{Blackley2014},
and dimensionless $\alpha$ and $\beta$ found from a fit to $V_{\rm q}$,
$g_{\rm q}$, and $a_{\rm q}$ in Eq.~\ref{eq:volumeFR}.  Lengths and
volumes are in units of $\bar a$ and $\bar a^3$, respectively, and 1 G=
0.1 mT. (Finally, $\bar a=42.95a_0$, $61.65a_0$, $43.63a_0$, and $96.51a_0$ for
$^{23}$Na, $^{39}$K, $^{52}$Cr, and $^{133}$Cs, respectively.)
 } \label{tab:res}

\begin{tabular}{c|c|c|c||c|c|c||c|c}
  &  $B_0$ &  $a_{\rm bg}$ & $s_{\rm res}$ & 
        $V_{\rm q}$ 
         &  $g_{\rm q}$ & $a_{\rm q}$
        & $\alpha$ & $\beta$ \\
\hline
$^{39}$K   & 745 &  -0.541 & 0.00062 & 4.7 & -1540 & -0.55 & 0.0354 & 0.0468\\
$^{133}$Cs &  227 &  21.34 & 0.19      & 1000 & -4.19 &  29 & -3.55 & -3.85\\
\hline
$^{23}$Na  &  853  &   1.47 & 0.0002    &  -  &   -     &   -     &    0   &  0 \\
$^{52}$Cr  & 500 & 2.45 & 0.03   &  -  &   -     &   -     &    0   &  0 \\
$^{133}$Cs & 19.8 &1.66 & 0.002 &  -  &   -     &   -     &    0   &  0
\end{tabular} 

\end{table}

\begin{figure}[b]
\begin{center}
	\includegraphics[width=0.45\textwidth,trim=0 10 30 20,clip]{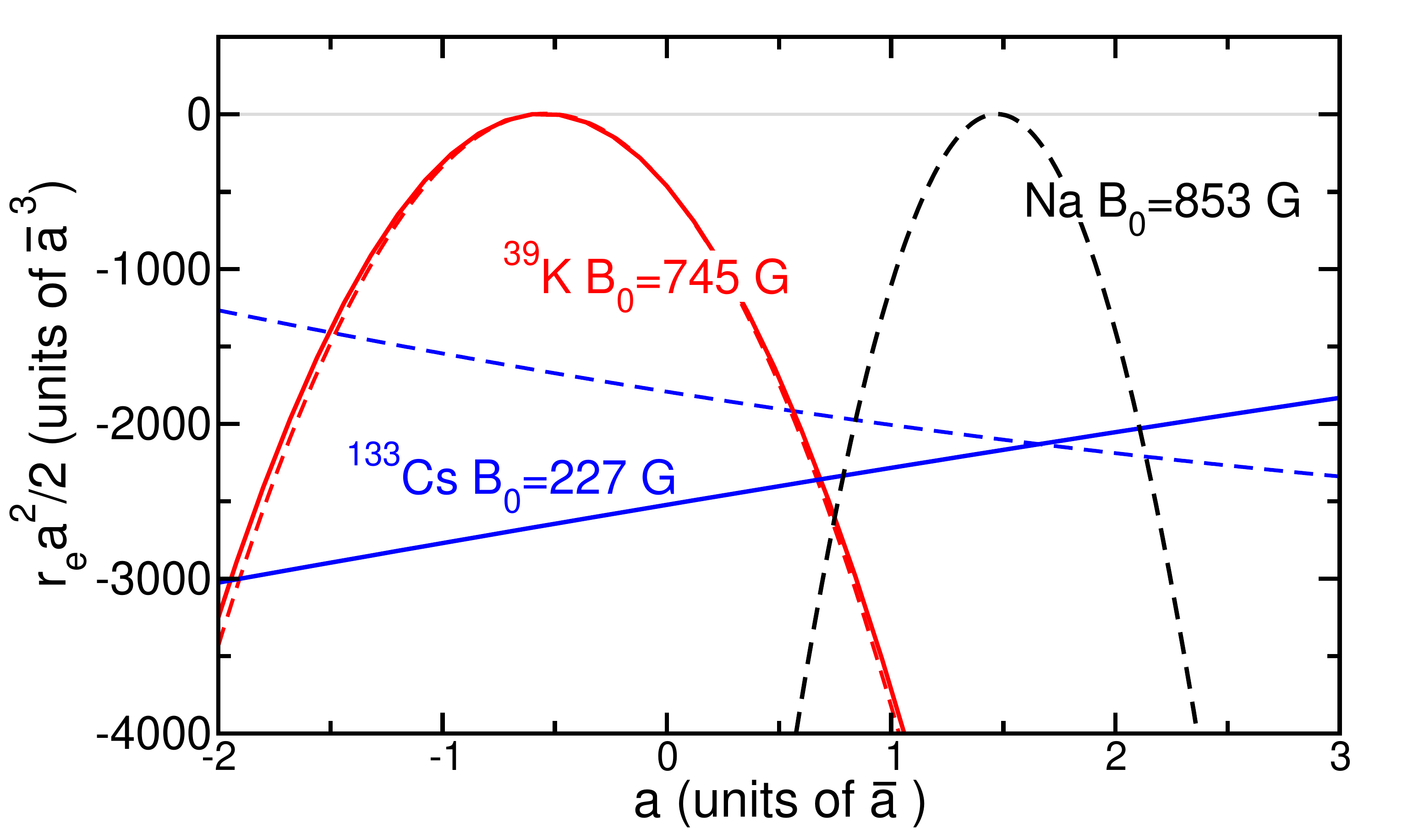}
\vspace*{-5mm}
\end{center}
\caption{(color online) Effective volume $r_ea^2/2$ of the Feshbach
resonances listed in Table \ref{tab:res} as a function of scattering
length $a$ with lengths in units of $\bar a$. 
Solid lines correspond to volumes based on 
Eq.~\ref{eq:direct} with $\alpha,\beta\ne0$
or equivalently the coupled-channels
calculation of \cite{Blackley2014}. Dashed lines follow
from Eq.~\ref{eq:direct} with $\alpha=\beta=0$.
} \label{fig:FB1}
\end{figure}

The effective-range volume near a resonance is a quadratic polynomial in
$a$ with coefficients defined by Eq.~\ref{eq:volumeFR}.  This dependence agrees
with the coupled-channels calculations with rigorous interatomic
potentials of Ref.~\cite{Blackley2014}.  Their $V_{\rm q}$, $g_{\rm q}$, and
$a_{\rm q}$ for a narrow $^{39}$K and broad Cs resonance are tabulated
in Table~\ref{tab:res}. The corresponding effective range volume as
well as that for a narrow Na resonance based on Eq.~\ref{eq:direct} with
$\alpha=\beta=0$ are shown in Fig.~\ref{fig:FB1} as a function of $a$ as
it is tuned with a magnetic field.  For narrow resonances $\alpha,\beta\ll
1$  and $\alpha,\beta$ have negligible effect on $r_ea^2/2$. For 
broad resonances with larger $\alpha,\beta$ their effect is large.
For both cases  $r_ea^2/2$ is negative and orders of magnitude
larger than that for van-der-Waals potentials.

The model for the effective range volume now enables us to
find scattering lengths where $U_2$ is small compared to $U_3$.
We set $U_2=0$ and Eq.~\ref{eq:cancelU2} gives
\begin{equation}
  V_{\rm q} + g_{\rm q} ( a- a_{\rm q} )^2 =  -\frac{2}{3}a\ell^2\,,
\end{equation}
where both $a$ and $\ell$ can be tuned. Coefficients $V_{\rm q}$,
$g_{\rm q}$, and $a_{\rm q}$ are fixed by the resonance.  Consequently,
choosing $a$ fixes the harmonic trapping frequency and vice versa.
Crucially and unlike for a van-der-Waals potential, $r_ea^2/2$ is
mostly negative and large compared to ${\bar a}^3$ so that $U_2=0$ can
occur for positive $a$ on the order of $\bar a$.  We must, however,
also require that $|r_ea^2/2|\ll \ell^3$.  This can not be guaranteed
for all resonances. For example, Fig.~\ref{fig:FB1} implies that for
the narrow $^{39}$K resonance and $a>{\bar a}$ the volume $|r_ea^2/2|\ge
\ell^3$ assuming typical $\ell$ between $ 10{\bar a}$ and $100{\bar a}$.
The narrow Na and broader Cs resonance show more promise.

\begin{figure}
\begin{center}
	\includegraphics[width=0.48\textwidth,trim=0 0 0 0,clip]{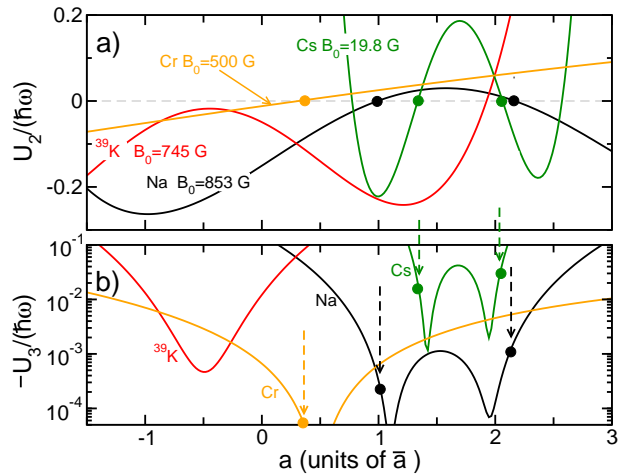}
\vspace*{-10mm}
\end{center}
\caption{(color online) Two-body interaction strength $U_2$ (panel a) and
minus one times the three-body strength $-U_3$ (panel b) in a harmonic
trap with $\omega/(2\pi)=50$ kHz as a function of the scattering length
$a$ for a narrow $^{23}$Na (black lines), $^{39}$K (red lines), $^{52}$Cr (orange
lines), and $^{133}$Cs (green lines) Feshbach resonance tabulated in
Table \ref{tab:res}. Filled circles in both panels and arrows in panel (b)
indicate where $U_2=0$.  }

\label{fig:U2}

\end{figure}

In Fig.~\ref{fig:U2} we make these observations more precise by plotting
$U_2$ and $U_3$ as a function of $a$ for four narrow Feshbach resonances
(with $s_{\rm res}\ll 0.1$) tabulated in Table \ref{tab:res} and assuming
a harmonic trap with frequency $\omega/(2\pi)=50$ kHz. For all four
resonances $U_2=0$ for at least one value of $a$. The second, broader
$^{133}$Cs resonance with $B_0=227$ G and $s_{\rm res}=0.19$ has
no such point and is not shown.  The cases where both $U_2=0$ and
$|U_3|/(\hbar\omega)\ll 1$ are indicated in the figure with markers.
For the Na and Cs resonance $U_2=0$ when $a\approx \bar a$ or $2\bar a$
and $-U_3/(\hbar\omega)\ge 0.001$.  For the $^{39}$K resonance
a zero crossing occurs at $a\approx 2 {\bar a}$ 
but $U_3/(\hbar\omega)\gg 1$, outside the validity range of
the theory. 

Finally, we compare the expected value of $U_3$ with the tunneling energy
$J$, depicted in Fig.~\ref{fig:zpeJU}, in an optical lattice. Noting
that for commonly-used lasers in and near the optical domain the recoil
energy $E_R/h$ lies between 2 kHz and 10 kHz for alkali-metal atoms,
we find that for $\omega/(2\pi)=50$ kHz the tunneling energy $J$ is
about ten times smaller than $|U_3|$. For a shallower lattice and thus
smaller $\omega$ the tunneling energy increases exponentially, while $U_3$,
maintaining the condition that $U_2=0$, decreases much more slowly.

\section{Detection and three-body recombination} \label{sec:recomb}

Several observations can be made about the feasibility and
limitations of the proposal. These range from the detection of the
point where $U_2=0$, the behavior of Bose-Hubbard models, and three-body
recombination.  The next two subsections will briefly address these
points.

\subsection{Detection of $U_2=0$?}

We can locate lattice parameters where $U_2=0$  with two types of experiments.
The simplest is to perform vibrational spectroscopy on two
or three isolated bosonic atoms held in a dipole trap or in an optical
lattice where tunneling is negligible.  For pairs of fermionic alkali-metal
atoms as well as for one fermion and one boson in an optical
lattice site this has been shown to work near a Feshbach resonance
by Refs.~\cite{Stoferle2006,Ospelkaus2006}.  Based on predictions of
\cite{Tiesinga2000} they found a new class of confinement-induced bound
states for large scattering lengths.  An accurate study for smaller
scattering lengths on the order of the mean scattering length or less,
however, is lacking for both fermionic and bosonic alkali-metal atoms.
For $^{88}$Sr with its small, negative scattering no such measurements
have been performed. Finally, no spectroscopic experiments for three-atoms
exist.

A second type of experiments that can locate $U_2=0$ are so-called
collapse-and-revival experiments in optical lattices, where changes of
the lattice parameters induce non-equilibrium dynamics. Specifically, realizations where 
after a sudden and large increase of the lattice depth
tunneling is negligible, the values for $U_2$ and $U_3$ can be inferred
from measurements of the momentum distribution as a function of delay after
the ramp \cite{Greiner2002b,Sebby-Strabley2007,Johnson2009,Will2010}.
In these experiments the initial state is a superfluid and, hence,
to good approximation each site contains a superposition of atomic
Fock states in the lowest trap level. After the sudden lattice-depth
increase this superposition starts to evolve and measurement of the
momentum distribution is sensitive to differences of the energies $E_n$
for different $n$.  These measurements have not been repeated near
Feshbach resonances.

\subsection{Three-body recombination} 

Atom loss from the lattice can limit the realization of our proposal.
Loss of one atom at a time, due to collisions with background molecules
in the vacuum or light-induced loss from the lattice lasers, can be
mitigated by improving the vacuum pressure and a careful choice of laser
frequencies.  Two-body loss can always be removed by choosing 
the hyperfine state with the lowest internal energy.  This leaves inelastic
three-body recombination as an intrinsic loss mechanism.  An ultra-cold
homogeneous thermal gas with number density $n$ loses atoms according
to rate equation $dn/dt=-3K_3 n^3$.  For scattering from short-range
potentials \cite{Braaten2007,Esry1999} the event rate coefficient $K_3\le
C_{\rm max}\hbar a^4/m$ with $C_{\rm max}=67$ when the scattering length
$|a|\gg \bar a$, while $K_3\approx C_0 \hbar {\bar a}^4/m$ with $C_0=25$
for $|a|\sim \bar a$.  Recently, Refs.~\cite{Wang2012,Wang2014} showed that
for longer-ranged van-der-Waals potentials and near collisional resonances
$C_0$ depends on atomic species and resonance, and can be much larger than
25.  Finally, Ref.~\cite{Shotan2014} showed empirically that for a broad
$^7$Li resonance with a negative effective range $K_3\approx  C_{\rm max}
\hbar (a^3-r_ea^2/2)^{4/3}/m$ gives a reasonable description of
experimental data close to the resonance.

In a lattice site recombination can be included as an imaginary
contribution to $U_3$. That is we use $U_3\to U_3-i\Gamma_3/2$, where $\Gamma_3 = \hbar
K_3 \int d^3\vec x |\Psi(\vec x)|^6$ and $\Psi(\vec x)$ is the normalized single-atom
ground-state wavefunction in a lattice site.  For an isotropic harmonic
trap and $|a|\sim \bar a$ this leads to
\begin{equation}
        \Gamma_3 =
            \frac{C_0}{3\pi^3} \frac{{\bar a}^4}{\ell^4} \, \hbar\omega
\end{equation}
when $U_2=0$.
Losses are acceptable when $\Gamma_3\ll |U_3|$ and  thus
\begin{equation}
            \frac{|a|}{\bar a } \gg  \sqrt{\frac{3\sqrt{3}C_0}{32\pi^2}} \frac{\bar a}{\ell}
 =0.64 \frac{\bar a}{\ell} 
\end{equation}
for $C_0=25$.
Since typically $\ell > 10 {\bar a}$, a scattering length on the order of
$\bar a$ is required. This condition can be met with Feshbach resonances,
but also indicates that an experiment with $^{88}$Sr will be hard.
A similar analysis with more restrictive estimate of Ref.~\cite{Shotan2014} 
suggests that weaker trapping potentials with $\ell \gg 10 {\bar a}$
will be required.

\section{Conclusion}

We have proposed a means to create an ultra-cold gas of bosonic atoms
in an optical lattice that only interacts via on-site three-body
interactions. This is achieved by a careful cancellation of two
contributions in the pair-wise interaction between two atoms,
one proportional to the zero-energy scattering length and a second
proportional to the effective range. We predict that this cancellation
can occur for the strontium-88 isotope as well as near narrow magnetic
Feshbach resonances in alkali-metal atom or chromium collisions.

For optical lattice depths and/or magnetic field strengths where the
pair-wise interaction has been cancelled, i.e. $U_2=0$, we have also
shown that the three-body interaction strength can be of the same order
of magnitude as the tunneling energy of atoms hopping between neighboring
lattice sites.  Three-body recombination can limit the practical duration
of coherent atom evolution.

Although the purpose of this paper has not been the characterization of the
many-body ground state or the dynamical properties of a system near
$U_2=0$, a brief remark is in order. For a small number
of atoms per lattice site we predict that the three-body interaction is
attractive. For a Hubbard model with finite tunneling $J$ on the
order of $U_3$ this can indicate that the ground state corresponds to
a state with all atoms in one site and, in essence, the system would
``collapse'', similar to the instability of systems with a negative
two-body strength $U_2$. To prevent this collapse a weak global trapping
potential must be added.  On the other hand, we expect that it is
realistic to perform dynamical experiments where initially the ground
state for positive $U_2$ is prepared and, subsequently, the lattice
parameters are changed to ones where $U_2=0$.

\section{Acknowledgments}

This work has been supported by the National Science Foundation Grant No. PHY-1506343.

\bibliography{refs}
\end{document}